# BaAs$_3$: A narrow gap 2D semiconductor with vacancy-induced semiconductor-metal transition

Ping Tang, [1+] Jun-Hui Yuan, [1+] Ya-Qian Song, [1] Ming Xu, [1] Kan-Hao Xue, [1,2*] and Xiang-Shui Miao[1]

[1]Wuhan National Research Center for Optoelectronics, School of Optical and Electronic Information, Huazhong University of Science and Technology, Wuhan 430074, China

[2]Univ. Grenoble Alpes, Univ. Savoie Mont Blanc, CNRS, Grenoble INP, IMEP-LAHC, 38000 Grenoble, France

[+]The authors P. Tang and J.-H. Yuan contributed equally to this work.

**Corresponding Author**

*E-mail: xkh@hust.edu.cn (K.-H. Xue)

**ABSTRACT:** Searching for novel two-dimensional (2D) materials is highly desired in the field of nanoelectronics. We here propose a new 2D crystal barium tri-arsenide (BaAs$_3$) with a series of encouraging functionalities. Being kinetically and thermally stable, the monolayer and bilayer forms of BaAs$_3$ possess narrow indirect band gaps of 0.87 eV and 0.40 eV, respectively, with high hole mobilities on the order of ~10$^3$ cm$^2$ V$^{-1}$ s$^{-1}$. The electronic properties of 2D BaAs$_3$ can be manipulated by controlling the layer thickness. The favorable cleavage energy reveals that layered BaAs$_3$ can be produced as a freestanding 2D material. Furthermore, by introducing vacancy defects monolayer BaAs$_3$ can be transformed from a semiconductor to a metal. 2D BaAs$_3$ may find promising applications in nanoelectronic devices.

Two-dimensional (2D) materials, such as graphene, transitional metal dichalcogenides (TMDCs) and phosphorene, hold great application potential for nanoelectronics, optoelectronics, energy storage and catalysis [1–12]. For example, monolayer black phosphorus or phosphorene, with a desired direct band gap of 2.0 eV, as well as a high hole mobility of about $1.14\times10^3$ cm$^2$ V$^{-1}$ s$^{-1}$-$2.60\times10^4$ cm$^2$ V$^{-1}$ s$^{-1}$, has been considered a strong candidate for next generation high performance field effect transistors [10,13]. However, the weak chemical stability under ambient conditions and the low electron mobility still hinder its practical application. Nowadays, searching for novel 2D materials with unique structural and desired electronic properties is still an active area of research.

Very recently, a series of 2D pnictides (nitrides, phosphides and arsenides) semiconductors have been theoretically proposed as novel 2D semiconductors with high carrier mobilities (~$10^3$ cm$^2$ V$^{-1}$ s$^{-1}$-$10^5$ cm$^2$ V$^{-1}$ s$^{-1}$) that are comparable or superior to those of phosphorene [14–25]. For instance, in monolayer Pt$_2$N$_4$ with a planar penta-structure, the electron mobility at room temperature can reach $1.1\times10^5$ cm$^2$ V$^{-1}$ s$^{-1}$, comparable to that in graphene [20]. In addition to the materials with a single atomic layer thickness, such as TM$_2$X$_4$(TM=Ni, Pd and Pt, X=P, As) [23], most of them possess blue phosphorene-type or black phosphorene-type structural features. For example, CaP$_3$ (CaAs$_3$) monolayers possess similar 2D networks of puckered configurations, and the puckered polyanionic $P_3^{2-}$ ($As_3^{2-}$) nets are derivatives of the black phosphorene (puckered arsenene) structure by removing 1/4 of the P (As) atoms. Moreover, bulk barium triarsenide ($BaAs_3$), also a member of $CaP_3$ families as $CaAs_3$, $SrP_3$ and $SrAs_3$, was first synthesized in 1981 [26]. Although known for decades, the research on $BaAs_3$ has been very rare. Recently, the bulk $CaP_3$ family has been predicted to be topological nodal-line semimetals [27], which may open a new research aspect for these materials. Compared with $CaP_3$ and $CaAs_3$, which belong to a low symmetry (*P*-1 or *C*-1), $BaAs_3$ has a higher space group symmetry *C*2/*m*. The stacking structural feature of $BaAs_3$ is similar to $CaP_3$ or $CaAs_3$, while $CaP_3$ and $CaAs_3$ were predicted to be stable when attenuated to atomic layer thickness, and exhibiting extraordinary optoelectronic properties. Thus, two questions are naturally raised. (i) Is $BaAs_3$ film of the atomic layer thickness stable? (ii) Given that few-layer $BaAs_3$ can be fabricated as free-standing 2D material, what about its electronic structure?

To answer these questions, in this work we have systematically investigated the stability and electronic properties of bulk $BaAs_3$ and freestanding monolayer $BaAs_3$. First, the kinetic and thermodynamic stabilities of monolayer $BaAs_3$ are confirmed according to phonon dispersion and high temperature molecular dynamics simulation. Then, we shall investigate the band gaps, carrier mobilities, and optical absorption characteristics of monolayer and few-layer $BaAs_3$. In particular, a detailed study of the vacancy defects in monolayer $BaAs_3$ with various concentrations have been carried out by employing the self-energy corrected shell GGA-1/2 method.

**Computational Methods**

We performed density functional theory (DFT) calculations using plane-wave-based Vienna *Ab-initio* Simulation Package (VASP) [28,29]. The generalized gradient approximation (GGA) within the Perdew-Burke-Ernzerhof (PBE) [30] functional form was used for the exchange-correlation energy, and projector augmented-wave pseudopotentials [31,32] were used to replace the core electrons. The valence electron configurations of Ba and As are 5$s$, 5$p$, 6$s$ and 4$s$, 4$p$, respectively. The screened exchange hybrid functional HSE06 [33] was used to calculate the electronic band structures in order to rectify the band gaps in GGA-PBE. The plane wave kinetic energy cutoff was fixed to be 500 eV. The van der Waals interactions were corrected by the DFT-D3 approach [34]. During structural relaxations, the convergence criterion for total energy was set to $1.0 \times 10^{-6}$ eV, and structural optimization was obtained until the Hellmann-Feynman force acting on any atom was less than 0.01 eV/Å in each direction. The phonon dispersion was calculated with the density functional perturbation theory, using the PHONOPY code [35]. *Ab initio* molecular dynamics (AIMD) simulations were performed to examine the thermal stability of the structures, where NVT canonical ensembles were used [36].

While the screened exchange hybrid functional HSE06 is well-known for its high quality electronic structure results, the slow convergence of the Hartree-Fock part yields much higher computational load than conventional GGA. In order to investigate the electronic structures of defects, where large supercells are used, we also implemented the shell GGA-1/2 method (shGGA-1/2) for self-energy correction [37].

The method is a variant of the original GGA-1/2 method proposed by Ferreira and coworkers in 2008 [38], with focus on better treatment of the covalent bonding. Self-energy correction was carried out on the As anions, where the inner and outer cutoff radii for the As self-energy potential [39] were fixed in a variational way, where the optimal set of cutoff radii should maximize the band gap.

**Results and discussion**

*Geometric structures*

The symmetry of bulk barium triarsenide ($BaAs_3$) is monoclinic with space group C2/*m*, with the optimized structure shown in **Fig.1a-c**. Bulk $BaAs_3$ has a 2D network of puckered configurations in-plane, and different van der Waals layers stacking out-of-plane. The puckered polyanionic $As_3^{2-}$ nets derive from the grey arsenene structure, with a quarter of As atoms missing, similar to $CaAs_3$ [19]. Our optimized lattice parameters of bulk $BaAs_3$ obtained using the D3-Grimme correction are $a$ = 10.22 Å, $b$= 7.82 Å, $c$ = 6.06 Å, and β=113.70°, which is in good accordance with the experimental results ($a$ = 11.16 Å, $b$= 7.76 Å, $c$ = 6.01 Å and β=113.55°) [26]. We also examined the electronic properties of bulk $BaAs_3$ based on GGA-PBE and HSE06 calculations, with or without considering the effect of spin-orbit coupling (SOC). As shown in **Fig. 1d**, regardless of the functional used, bulk $BaAs_3$ is a semimetal when SOC is neglected. In addition, the semimetallic nature is retained when SOC is considered (see band structures in **Fig. S1**).

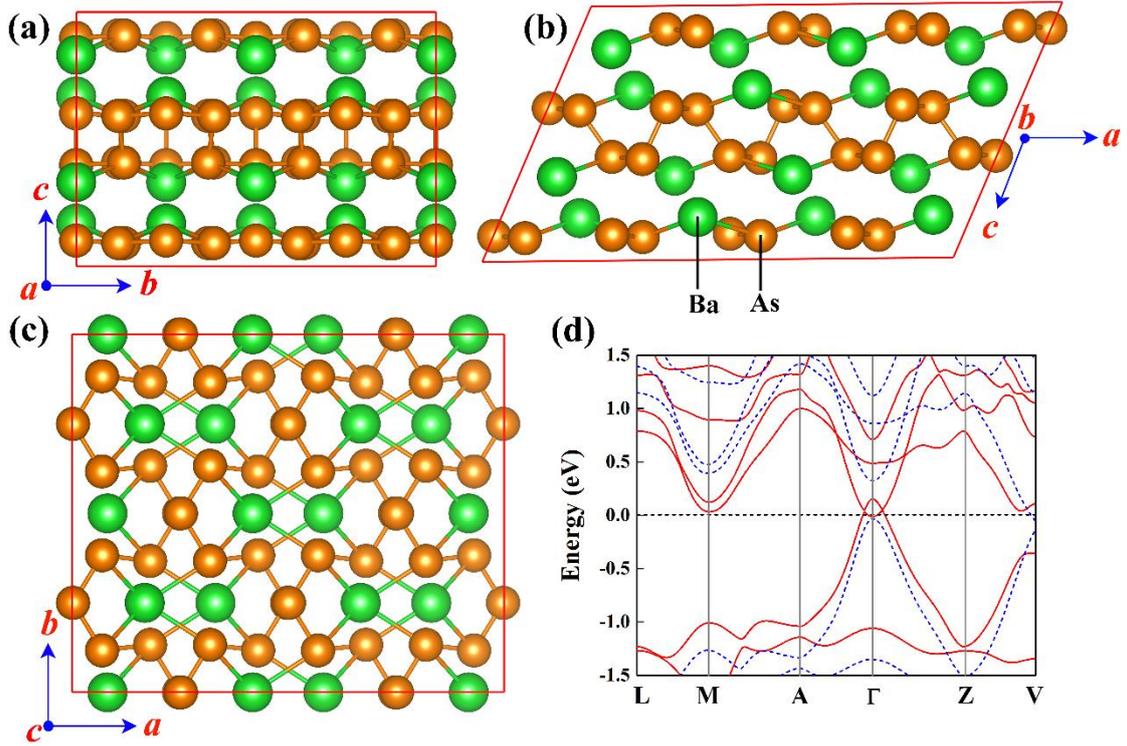

**Figure 1** (a,b,c) Crystal structure of optimized bulk BaAs$_3$, showing a 2×2×2 supercell from three distinct perspectives. The green and brown ball represent Ba and As atoms, respectively. (d) Calculated electronic band structures of bulk BaAs$_3$ using GGA-PBE (solid red line) and HSE06 (dash blue line), respectively. The Fermi levels are set to zero energy.

When exfoliating from the bulk, monolayer (ML for short) BaAs$_3$ has in theory the same geometric structures with high stability, as plotted in **Fig. 2a**. The optimized lattice parameters of ML BaAs$_3$ are $a$ =6.51 Å, $b$ = 6.25 Å and γ=73.19°. The As-As bond length ranges from 2.48 Å to 2.53 Å, similar to that of puckered arsenene (2.485 Å to 2.501 Å) [40], while the Ba-As bond length is from 3.24 Å to 3.82 Å. Compared with the bulk phase, however, a slight distortion for As-As bond length (2.48 Å to 2.49 Å for bulk) and Ba-As bond length (3.29 Å to 3.79 Å for bulk) have been found in the ML form. Hence, careful theoretical verification is still required for the stability of the proposed ML BaAs$_3$ structure.

First of all, we evaluate the feasibility of exfoliating the ML BaAs$_3$ sheet from its layered bulk crystal, from a cleavage energy aspect. A five-layer BaAs$_3$ slab is utilized to server as a model of the bulk. The computed cleavage energy is 1.02 J m$^{-2}$, as illustrated in **Fig. 2c**. The DFT-estimated exfoliation energy of ML BaAs$_3$ is larger than that of graphite (0.37 J m$^{-2}$ from experimental and 0.32 J m$^{-2}$ in theory) [41,42] but at the same level of InP$_3$ (1.32 J m$^{-2}$) [14], GeP$_3$ (1.14 J m$^{-2}$) [15], CaP$_3$ (1.30 J m$^{-2}$) [18] and CaAs$_3$ (1.36 J m$^{-2}$) [19]. Therefore, in principle ML BaAs$_3$ crystal could be prepared experimentally from its bulk counterpart using mechanical cleavage or liquid phase exfoliation. Subsequently, we focus on the kinetic stability and thermal stability, which are crucial for real experimental fabrication. The kinetic stability was assessed by calculating the phonon dispersion relations. As shown in **Fig. 2d**, no imaginary phonon modes are identified, indicating that ML BaAs$_3$ is kinetically stable. The highest phonon mode of ML BaAs$_3$ is 237 cm$^{-1}$, which is comparable to that of puckered arsenene (253 cm$^{-1}$) [40]. In addition, the mode stems from the As-As bond, as revealed by the phonon density of states results (**Fig. 2d**). Furthermore, the thermodynamic stability of ML BaAs$_3$ was assessed by performing AIMD simulations. As pointed out by the structural snapshots and variations of total energy in **Fig. S2**, ML BaAs$_3$ maintains its structure up to 500 K within 10 ps, with no geometric reconstruction or bond breaking discovered during the whole process. Combining the results from cleavage energy, phonon dispersion as well as high temperature AIMD, we conclude that ML BaAs$_3$ can be exfoliated from the bulk crystal, staying as a freestanding 2D material.

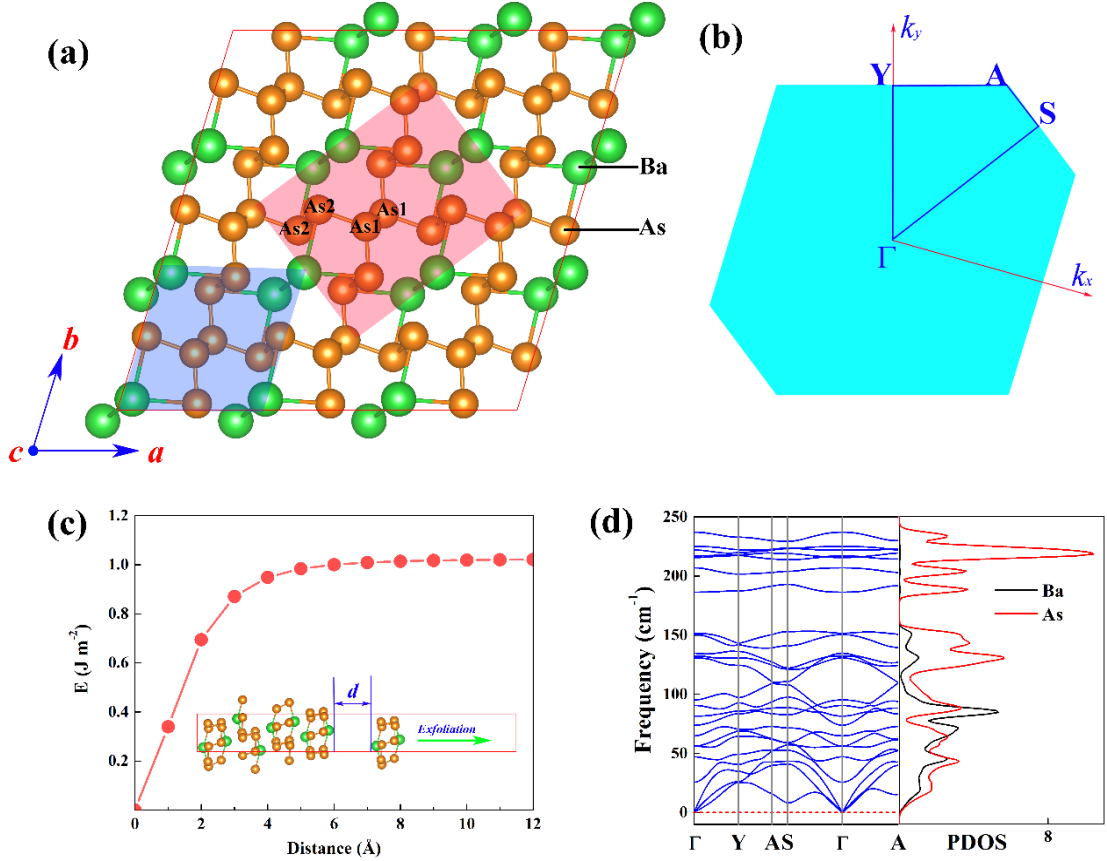

**Figure 2** (a) Top view of the optimized ML BaAs$_3$ in a 3×3 supercell. The light blue and light red areas denote the primitive cell and the corresponding rectangular cell of ML BaAs$_3$, respectively. (b) The corresponding first Brillouin zone of ML BaAs$_3$ primitive cell with high symmetry points marked. (c) Cleavage energy estimation for the formation of ML BaAs$_3$, calculated by enlarging the interlayer distance between the ML system that is removed from the remainder of a five-layer slab. (d) Calculated phonon dispersion spectra and phonon density of states of the ML BaAs$_3$.

*Electronic properties*

To understand the bonding characteristics of ML BaAs$_3$, we calculated its electron localization function (ELF) [43,44] and Bader charges [45–47]. As shown in **Fig. S3**, the ionic and covalent characters for Ba-As and As-As bonds, respectively, are clearly demonstrated through ELF analysis. Interestingly, ELF and Bader charge analysis (results summarized in **Table S1**) reveal there are two different types of As in BaAs$_3$, As1 connected by As-As bonds only and As2 connected by both Ba-As and As-As bonds, respectively. About 1.32|e| charge on average has been transferred from the Ba atom to the neighboring As2 atoms (∼0.56-0.57 |e|), while the As1 atom in the As−As bond gains a low amount of charge (∼0.18 |e|). The As-As bond in ML BaAs$_3$ is much different from that of puckered arsenene, whose As atoms in the As-As bonds are almost all neutral (∼0.025/0.033 |e|), as shown in **Fig. S4b**. In sum, the hybrid ionic and covalent bonds between Ba and As atoms are jointly responsible for the formation and stability of the ML BaAs$_3$.

Subsequently, we focus on the energy band structure of ML BaAs$_3$. Heavy elements like As may render strong spin-orbit coupling (SOC) effect. Hence, we calculate band structures of ML BaAs$_3$ both with and without considering SOC, using either the PBE functional or the HSE06 hybrid functional. Meanwhile, the new shGGA-1/2 self-energy correction method is also used in band structure calculations for comparison. As shown in **Figure 3a**, using GGA-PBE without turning on SOC, the ML BaAs$_3$ is predicted to be an indirect semiconductor with a narrow band gap value of 0.16 eV. The valence band maximum (VBM) is located at the $\Gamma$-point while the conduction band minimum (CBM) lies along the $S$-$\Gamma$ direction, closer to the $S$ point. Similar band structures but with larger band gap values of 0.75 eV/0.87 eV have been confirmed based on HSE06/shGGA-1/2 calculations, respectively. We then examine the effect of SOC in our calculations. The CBM and VBM of ML BaAs$_3$ do not show discernable shift after turning on the SOC, regardless of using PBE or HSE06, as plotted in **Fig.S5**. The SOC effect produces less than 0.01 eV variation in the band gap. Therefore, we shall neglect the SOC effect in all forthcoming band structure calculations. Furthermore, the partial density of states (PDOS) results show that the As-4$p$ (especially the As2-4$p$) states dominate the orbit contribution around the Fermi level, which is further confirmed by the spatial charge distributions of VBM and CBM (shown in **Figs. 3b and 3c**).

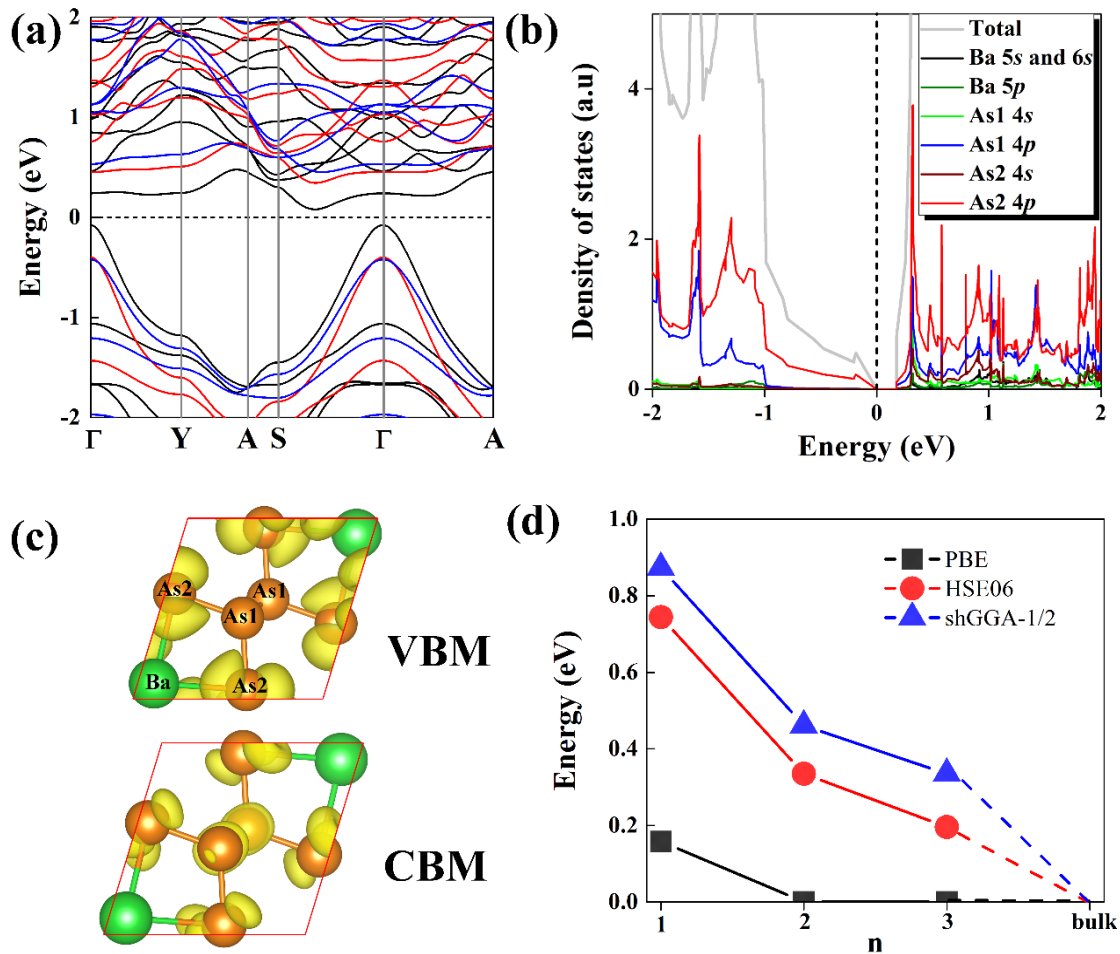

**Figure 3** (a) Electronic band structures of ML BaAs$_3$ calculated using GGA-PBE (solid black line), HSE06 (solid red line) and shGGA-1/2 (solid blue line) without considering SOC. (b) Partial DOS of the ML BaAs$_3$ calculated using the PBE functional. (c) Spatial distribution of the wave-functions corresponding to the VBM and CBM of ML BaAs$_3$ (contour density: 0.02 e Å$^{-3}$). (d) Computed band gaps of BaAs$_3$ multilayers versus the number of atomic layers, using GGA-PBE, HSE06 hybrid functional and shGGA-1/2, respectively.

The ML BaAs$_3$ owns a narrow band gap while the bulk BaAs$_3$ is calculated to be a semimetal. In order to elucidate the changes in the electronic properties of BaAs$_3$ from the bulk to few layers, we have investigated the electronic band gaps of 2D BaAs$_3$ with varying number of layers, with the results shown in **Fig. 3d** and **Fig. S6**. The electronic structures of 2D BaAs$_3$ multilayers indeed strongly depend on the number of layers. At the PBE level, ML BaAs$_3$ is the only semiconductor while bilayer (BL for short) and tri-layer (TL for short) BaAs$_3$ remain metallic. According to both HSE06 and shGGA-1/2 calculations, 2D BaAs$_3$ up to 3 layers can still maintain the indirect band gap feature (**Fig.S6**), though the band gap is diminished for more layers. For BL BaAs$_3$, the indirect band gap is 0.33 eV/0.40 eV (HSE06/shGGA-1/2), while the TL BaAs$_3$ possesses an indirect band gap of 0.20 eV/0.34 eV (HSE06/shGGA-1/2). Therefore, the band gap of BaAs$_3$ can be engineered by controlling the layer thickness to cover a relatively large range of 0-0.75 eV/0-0.87 eV (HSE06/shGGA-1/2). Detailed optimized lattice constants of BL, TL and bulk BaAs$_3$ are shown in **Table S2**.

*Carrier Mobility*

The ML BaAs$_3$ exhibits a narrow band gap close to that of germanium (0.74 eV)[37], suggesting it may find potential application in electronic devices. Therefore, we systematically calculated the carrier mobilities (for electrons and holes) of ML and BL BaAs$_3$ based on the deformation potential theory of Bardeen and Shockley [48]. For 2D semiconductors, Lang *et al*. [49] recently gave a new formula for the acoustic phonon-limited mobility, taking into account the anisotropic elastic constants or deformation potential constants. For nonisotropic 2D materials, the theoretical carrier mobility can be calculated by the following equation [42,49,50]

$$\mu_{2D} = \frac{e\hbar^3 \left( \frac{5C_a^{2D} + 3C_b^{2D}}{8} \right)}{k_B T (m_a)^{\frac{3}{2}} (m_b)^{\frac{1}{2}} (\frac{9E_{1a}^2 + 7E_{1a}E_{1b} + 4E_{1b}^2}{20})^2},$$

where $\hbar$ is the reduced Planck constant, $k_B$ is the Boltzmann constant, $m^*$ is the effective mass in the direction of transport, $m_d$ is the average effective mass determined by $m_d = (m_a^* m_b^*)^{1/2}$, and $T$ is the temperature ($T$ = 300 K). The elastic modulus $C_{2D}$ of the longitudinal strain in the propagation direction is derived from $(E - E_0)/S_0 = C_{2D}(\Delta l / l_0)^2 / 2$, where $E$ is the total energy of the 2D structure, and $S_0$ is the lattice area of the equilibrium supercell. The deformation potential constant $E_1^i$ is defined as $E_1^i = \Delta E_i / (\Delta l / l_0)$. Here $\Delta E_i$ is the energy change of the $i^{th}$ band under proper cell compression and dilatation (calculated using a step of 0.5%), $l_0$ is the lattice constant in the transport direction and $\Delta l$ is the deformation of $l_0$. Notice that the directions *a* and *b* for the unit cell of ML BaAs$_3$ are not perpendicular to each other. Indeed, the γ angle is 73.19° that could not simply be approximated as 90°. Therefore, for ML BaAs$_3$ we adopt an orthogonal supercell (shown in **Fig.2a**) to calculate the carrier mobility. In this case, we have to re-calculate the corresponding energy band structures based on the orthorhombic supercell, for the sake of deformation potential constant evaluation. As shown in **Fig.S7**, the obtained band gap value and indirect gap feature using the orthorhombic supercell are in line with the results from the primitive cell. In addition, considering that GGA-PBE calculations do not recover the correct semiconducting nature for BL-BaAs$_3$, all the following energy band structures are calculated by HSE06 functional (except for elastic modulus, which can be obtained accurately at the PBE level).

**Table 1** Calculated effective mass $m^*$ ($m_e$, HSE06 results), deformation potential constant $|E_1^i|$ (eV, HSE06 results), elastic modulus $C_{2D}$ (N m$^{-1}$, PBE results), and carrier mobility $\mu_{2D}$ (cm$^2$ V$^{-1}$ s$^{-1}$) for ML and BL BaAs$_3$ along the *a* and *b* directions.

| Layers | Carrier type | $m_a^*$ | $m_b^*$ | $|E_{1a}|$ | $|E_{1b}|$ | $C_a^{2D}$ | $C_b^{2D}$ | $\mu_a^{2D}$ | $\mu_b^{2D}$ |
|---|---|---|---|---|---|---|---|---|---|
| ML | electron | 0.679 | 0.634 | 1.415 | 1.451 | 24.86 | 29.52 | 629.361 | 580.419 |
|    | hole     | 0.108 | 0.262 | 3.872 | 6.768 | 24.86 | 29.52 | 1030.549 | 369.646 |
| BL | electron | 0.736 | 0.953 | 2.314 | 3.779 | 60.14 | 60.14 | 249.554 | 151.994 |
|    | hole     | 0.121 | 0.298 | 4.299 | 9.00  | 60.14 | 60.14 | 1464.680 | 421.586 |

As summarized in **Table 1,** the carrier effective masses for ML BaAs$_3$ along the *a*/*b* direction are 0.679/0.634 $m_e$ for the electron and 0108/0.262 $m_e$ for the hole ($m_e$ is the free electron mass), while those for BL BaAs$_3$ are 0.736/0.953 $m_e$ and 0.121/0.298 $m_e$, respectively. The hole effect masses are much smaller than that of electrons, which can be expected from the more steep band structures around VBM. The elastic moduli are 24.86/29.52 N m$^{-1}$ and 60.14/60.14 N m$^{-1}$ for ML and BL BaAs$_3$ along the *a*/*b* directions, respectively. The deformation potential constants $E_1$ for the electron of ML and BL BaAs$_3$ are both smaller than that of the hole. Based on the above obtained $m^*$, $C_{2D}$ and $E_l$ values, we estimated the acoustic phonon-limited mobilities of ML and BL BaAs$_3$ as 629.361/580.419 and 249.554/151.994 cm$^2$ V$^{-1}$ s$^{-1}$ along *a*/*b* directions, respectively. In contrast to the relatively weak anisotropy of electron mobilities, the hole mobilities show stronger anisotropy with the value of 1030.549/369.646 cm$^2$ V$^{-1}$ s$^{-1}$ and 1464.680/421.586 cm$^2$ V$^{-1}$ s$^{-1}$ along *a* and *b* directions for ML and BL BaAs$_3$, respectively. The large anisotropy of hole mobilities mainly stems from the large effective mass and deformation potential constant along the *b* direction.

*Vacancy induced semiconductor-metal transition*

Point defects play a crucial role in the properties of materials, especially in electronic and optoelectronic devices. Therefore, it is very important to study the effect of different types of vacancies with various concentrations in theory and experiment. However, it is difficult to directly study point defects by experiments, thus first-principles calculations appear to be an indispensable means.

For this purpose, we perform calculations with varying supercell sizes (2×2×1 to 7×7×1) for a fairly large set of defects. A single Ba/As vacancy is introduced in each supercell, where large supercells (288/392 atoms for 6×6×1/7×7×1 supercell) lead to low vacancy concentrations. As mentioned above, there are two different types of As atom in BaAs$_3$. As far as the anion vacancy is concerned, we considered two different As vacancies, $V_{As1}$ and $V_{As2}$ respectively, while only one Ba cation vacancy ($V_{Ba}$) was considered.

The defect formation energy, for the case of one neutral defect $D$ per supercell, can be defined as [51]

$$E_f[D] = \{E[D] + \mu[D]\} - E_P$$

where $E[D]$ and $E_P$ are the total energies of the supercell with the defect $D$ and the perfect stoichiometric supercell, while $\mu[D]$ represents the chemical potential of $D$. We chose the Ba and As chemical potentials as in their ground state solid metal (*b.c.c.* Ba) and (*R-3m* As). The relation between the formation energy and the size of the supercell is illustrated in **Figure 4(a).**

As the size of the supercell goes from $2\times2\times1$ to $7\times7\times1$, the formation energies of the three vacancies show a uniform trend of decreasing. $E_f[\text{Ba}]$ drops from 2.60 eV to 2.41 eV, while $E_f[\text{As1}]/E_f[\text{As2}]$ also drops from 1.48 eV/0.63 eV to 1.15 eV/0.15 eV. For the same supercell scale, $E_f[\text{Ba}]$ is always larger than $E_f[\text{As}]$. Among the two different types of $V_{\text{As}}$, $V_{\text{As1}}$ shows a quite smaller formation energy than $V_{\text{As1}}$. In $7\times7\times1$ supercell, $E_f[\text{As2}]$ is merely 0.15 eV. To sum up, in ML BaAs$_3$ $V_{\text{Ba}}$ is difficult to create, while $V_{\text{As2}}$ is the most favorable defect.

After evaluating the feasibility of removing a single atom from different sizes of ML BaAs$_3$, we then turn our attention to the electronic properties of the vacancy-containing systems. The test results do not show any spin-polarized ground state in these defective supercells, so we have disabled spin polarization. The shGGA-1/2 method is utilized for fast and accurate electronic structure calculation, with comparison to conventional GGA. For Brillouin zones, the 32-atom supercells are sampled with a $7 \times 7 \times 1$ Monkhorst-Pack k-point mesh, the 72-atom and 128-atom supercells with a 5 × 5 × 1 mesh, and the 200-atom, 288-atom, 392 atom supercells with a 3 × 3 × 1 mesh, respectively.

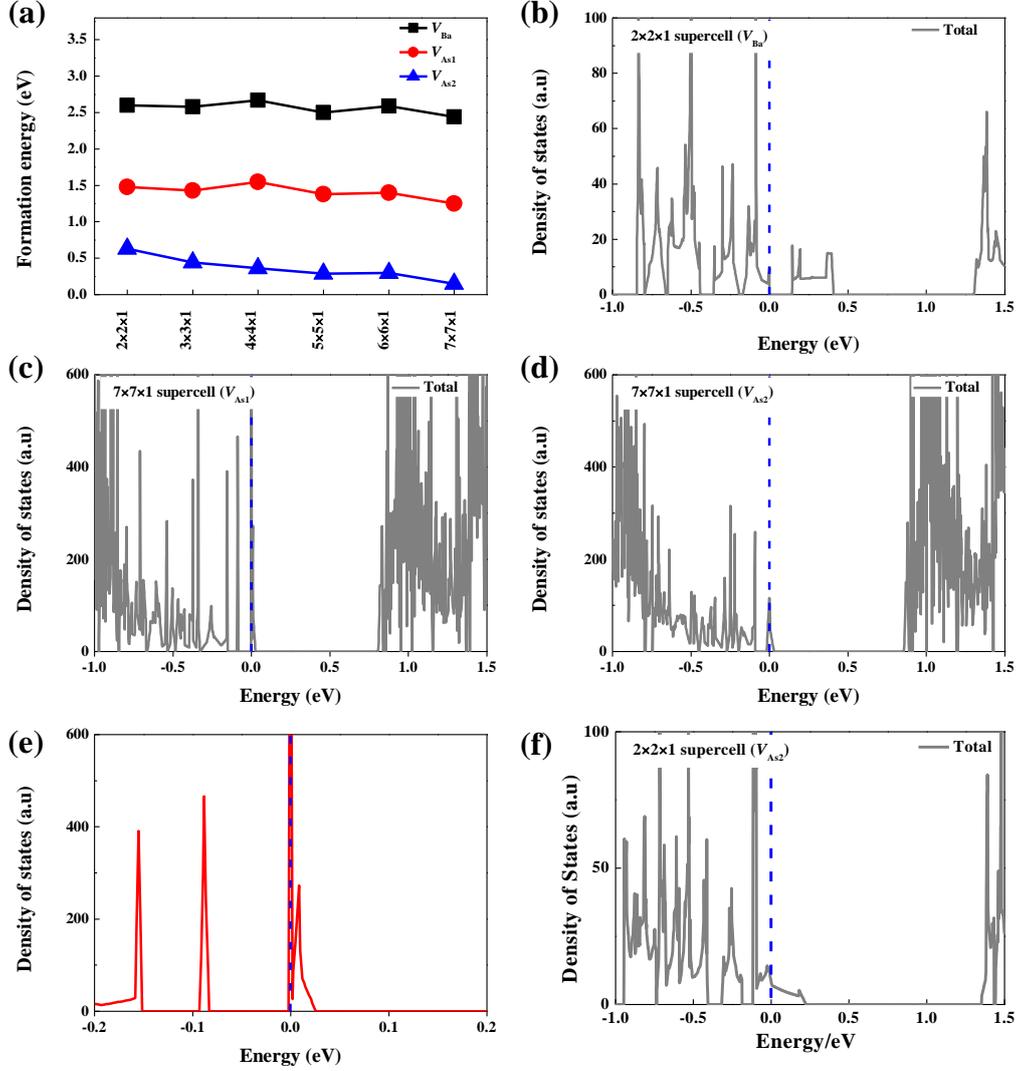

**Figure 4** (a) Relation between the defect formation energy and the size of the supercell, for Ba, 1-As and 2-As vacancies. Partial DOS of vacancy-containing supercells calculated using shGGA-1/2: (b) $V_{Ba}$ in a 2×2×1 supercell; (c) $V_{As1}$ in a 7×7×1 supercell; (d) $V_{As2}$ in a 7×7×1 supercell; (e) enlarged view of (c) near the Fermi level; (f) $V_{As2}$ in a 2×2×1 supercell.

With the vacancy concentration increases, the band gap of $V_{Ba}$-containing supercell gradually decreases **(Fig. S10)**. Yet, even at a very high concentration (3.12% in $2\times2\times1$ supercell), the cation-deficitient ML BaAs$_3$ still maintains a finite energy gap **(Figure4b)**. After introducing $V_{Ba}$ in a $2\times2\times1$ supercell, the band gap drops from 0.87 eV in perfect supercell to 0.14 eV (both using shGGA-1/2), which suggests that the conductivity of ML BaAs$_3$ can be adjusted by controlling the $V_{Ba}$ concentration in a quite large range.

Interestingly, the absence of As yields quite different results **(Fig. S11** and **Fig. S12)**. Both types of $V_{As}$ can transform ML BaAs$_3$ into a metal at an appropriate concentration. The DOS diagram shows that when the vacancy concentration is high, the density of states near the Fermi level is broadened. On the other hand, in the $7\times7\times1$ supercell a very sharp but narrow state appears **(Fig. 4c** and **Fig. 4d)**. The hole is highly localized with very tiny dispersion. The ML BaAs$_3$ can still be regarded as a semiconductor in this sense. Meanwhile, $V_{As2}$ is much more effective to trigger the semiconductor-metal transition, while it is in the meantime the most ordinary defect as well.

For most semiconductor materials, the conductivity of the material can be enhanced by n-type (typically anion vacancy) or p-type (typically cation vacancy) doping, both of which can convert the material to a metal after reaching a suitable concentration. However, for ML BaAs$_3$, the incorporation of all three kinds of vacancies only yields p-type conduction. This can be attributed to the unusual stoichiometry of the material, where the As vacancy actually does not act like a true anion vacancy (i.e., not inducing n-type conduction). This interesting finding is of great significance for the design of novel electronic devices in many fields, such as memristors.

## Conclusion

To summarize, we have predicted a novel 2D layered semiconductor $BaAs_3$ that various exceptional electronic properties. The monolayered $BaAs_3$ is predicted to be an indirect semiconductor with a quite narrow band gap of 0.87 eV and can be tuned by controlling the layer thickness to cover a relatively large range of 0.87 eV to 0 eV, accordingly to shell GGA-1/2 calculations. The possibility of exfoliating layer $BaAs_3$ from the bulk structure has been confirmed by the 1.02 J m$^{-2}$ cleavage energy. The kinetic and thermal stability are verified by the phonon spectrum as well as AIMD simulations, proving that layered $BaAs_3$ can be produced as a freestanding 2D material in experiment. Meanwhile, monolayer and bilayer $BaAs_3$ also show outstanding carrier mobility for holes (~$10^3$ cm$^2$ V$^{-1}$ s$^{-1}$). We find that the As2 vacancy is easy to form in monolayer $BaAs_3$, which surprisingly leads to p-type conduction. On the other hand, the Ba vacancy is not effective in causing electrical conduction in monolayer $BaAs_3$. The special electronic properties of monolayer $BaAs_3$ may be useful for novel nanoelectronic devices.

## ASSOCIATED CONTENT

## Conflicts of interest

There are no conflicts to declare.

# Acknowledgement

This work was supported by the National Key Research and Development Program of China (Materials Genome Initiative, 2017YFB0701700), the National Natural Science Foundation of China under Grant No. 11704134, and the Fundamental Research Funds of Wuhan City under Grant No. 2017010201010106. K.-H. Xue received support from China Scholarship Council (No. 201806165012).

# REFERENCES

[1] K.S. Novoselov, A.K. Geim, S.V. Morozov, D. Jiang, Y. Zhang, S.V. Dubonos, I.V. Grigorieva, A.A. Firsov, Electric Field Effect in Atomically Thin Carbon Films, Science. 306 (2004) 666–669. doi:10.1126/science.1102896.

[2] B. Radisavljevic, A. Radenovic, J. Brivio, V. Giacometti, A. Kis, Single-layer $MoS_2$ transistors, Nature Nanotechnology. 6 (2011) 147–150. doi:10.1038/nnano.2010.279.

[3] O. Lopez-Sanchez, D. Lembke, M. Kayci, A. Radenovic, A. Kis, Ultrasensitive photodetectors based on monolayer MoS2, Nature Nanotechnology. 8 (2013) 497–501. doi:10.1038/nnano.2013.100.

[4] C. Liu, Z. Yu, D. Neff, A. Zhamu, B.Z. Jang, Graphene-Based Supercapacitor with an Ultrahigh Energy Density, Nano Letters. 10 (2010) 4863–4868. doi:10.1021/nl102661q.

[5] M.D. Stoller, S. Park, Y. Zhu, J. An, R.S. Ruoff, Graphene-Based Ultracapacitors, Nano Letters. 8 (2008) 3498–3502. doi:10.1021/nl802558y.

[6] M.A. Lukowski, A.S. Daniel, F. Meng, A. Forticaux, L. Li, S. Jin, Enhanced Hydrogen Evolution Catalysis from Chemically Exfoliated Metallic $MoS_2$ Nanosheets, Journal of the American Chemical Society. 135 (2013) 10274–10277. doi:10.1021/ja404523s.

[7] D. Deng, K.S. Novoselov, Q. Fu, N. Zheng, Z. Tian, X. Bao, Catalysis with two-dimensional materials and their heterostructures, Nature Nanotechnology. 11 (2016) 218–230. doi:10.1038/nnano.2015.340.

[8] M.Z. Rahman, C.W. Kwong, K. Davey, S.Z. Qiao, 2D phosphorene as a water splitting photocatalyst: fundamentals to applications, Energy & Environmental Science. 9 (2016) 709–728. doi:10.1039/C5EE03732H.

[9] J. Ran, B. Zhu, S.-Z. Qiao, Phosphorene Co-catalyst Advancing Highly Efficient Visible-Light Photocatalytic Hydrogen Production, Angewandte Chemie International Edition. 56 (2017) 10373–10377. doi:10.1002/anie.201703827.

[10] L. Li, Y. Yu, G.J. Ye, Q. Ge, X. Ou, H. Wu, D. Feng, X.H. Chen, Y. Zhang, Black phosphorus field-effect transistors, Nature Nanotechnology. 9 (2014) 372–377.

doi:10.1038/nnano.2014.35.

[11] T. Fan, Y. Zhou, M. Qiu, H. Zhang, Black phosphorus: A novel nanoplatform with potential in the field of bio-photonic nanomedicine, Journal of Innovative Optical Health Sciences. (2018) 1830003. doi:10.1142/S1793545818300033.

[12] X. Zhang, L. Hou, A. Ciesielski, P. Samorì, 2D Materials Beyond Graphene for High-Performance Energy Storage Applications, Advanced Energy Materials. 6 (2016) 1600671. doi:10.1002/aenm.201600671.

[13] J. Qiao, X. Kong, Z.-X. Hu, F. Yang, W. Ji, High-mobility transport anisotropy and linear dichroism in few-layer black phosphorus, Nature Communications. 5 (2014) 4475. doi:10.1038/ncomms5475.

[14] N. Miao, B. Xu, N.C. Bristowe, J. Zhou, Z. Sun, Tunable Magnetism and Extraordinary Sunlight Absorbance in Indium Triphosphide Monolayer, J. Am. Chem. Soc. 139 (2017) 11125–11131. doi:10.1021/jacs.7b05133.

[15] Y. Jing, Y. Ma, Y. Li, T. Heine, $GeP_3$: A Small Indirect Band Gap 2D Crystal with High Carrier Mobility and Strong Interlayer Quantum Confinement, Nano Letters. 17 (2017) 1833–1838. doi:10.1021/acs.nanolett.6b05143.

[16] S. Sun, F. Meng, H. Wang, H. Wang, Y. Ni, Novel two-dimensional semiconductor $SnP_3$: high stability, tunable bandgaps and high carrier mobility explored using first-principles calculations, Journal of Materials Chemistry A. 6 (2018) 11890–11897. doi:10.1039/C8TA02494D.

[17] J.-H. Yuan, A. Cresti, K.-H. Xue, Y.-Q. Song, H.-L. Su, L.-H. Li, N.-H. Miao, Z.-M. Sun, J.-F. Wang, X.-S. Miao, $TlP_5$: an unexplored direct band gap 2D semiconductor with ultra-high carrier mobility, Journal of Materials Chemistry C. (2019) 639-644. doi:10.1039/C8TC05164J.

[18] N. Lu, Z. Zhuo, H. Guo, P. Wu, W. Fa, X. Wu, X.C. Zeng, $CaP_3$: A New Two-Dimensional Functional Material with Desirable Band Gap and Ultrahigh Carrier Mobility, The Journal of Physical Chemistry Letters. 9 (2018) 1728–1733. doi:10.1021/acs.jpclett.8b00595.

[19] F. Li, H. Wu, Z. Meng, R. Lu, Y. Pu, Tunable Topological State, High Hole-Carrier Mobility, and Prominent Sunlight Absorbance in Monolayered Calcium Triarsenide, The Journal of Physical Chemistry Letters. 10 (2019) 761–767. doi:10.1021/acs.jpclett.9b00033.

[20] Z. Liu, H. Wang, J. Sun, R. Sun, Z.F. Wang, J. Yang, Penta-$Pt_2N_4$: an ideal two-dimensional material for nanoelectronics, Nanoscale. 10 (2018) 16169–16177. doi:10.1039/C8NR05561K.

[21] J.-H. Yuan, Y.-Q. Song, Q. Chen, K.-H. Xue, X.-S. Miao, Single-layer planar penta-X2N4 (X= Ni, Pd and Pt) as direct-bandgap semiconductors from first principle calculations, Applied Surface Science. (2018) 456-462. doi:10.1016/j.apsusc.2018.11.041.

[22] H. Yuan, Z. Li, J. Yang, Atomically thin semiconducting penta-$PdP_2$ and $PdAs_2$ with ultrahigh carrier mobility, Journal of Materials Chemistry C. 6 (2018) 9055–9059. doi:10.1039/C8TC03368D.

[23] J.-H. Yuan, B. Zhang, Y.-Q. Song, J.-F. Wang, K.-H. Xue, X.-S. Miao, Planar penta-transition metal phosphide and arsenide as narrow-gap semiconductors with ultrahigh carrier mobility, Journal of Materials Science. 54 (2019) 7035–7047. doi:10.1007/s10853-019-03380-4.

[24] B. Ghosh, S. Puri, A. Agarwal, S. Bhowmick, $SnP_3$: A Previously Unexplored Two-Dimensional Material, The Journal of Physical Chemistry C. 122 (2018) 18185–18191. doi:10.1021/acs.jpcc.8b06668.

[25] S. Yao, X. Zhang, Z. Zhang, A. Chen, Z. Zhou, 2D Triphosphides: SbP3 and GaP3 monolayer as promising photocatalysts for water splitting, International Journal of Hydrogen Energy. 44 (2019) 5948–5954. doi:10.1016/j.ijhydene.2019.01.106.

[26] W. Bauhofer, M. Wittmann, H.G. v. Schnering, Structure, electrical and magnetic properties of CaAs3, SrAs3, BaAs3 and EuAs3, Journal of Physics and Chemistry of Solids. 42 (1981) 687–695. doi:10.1016/0022-3697(81)90122-0.

[27] Q. Xu, R. Yu, Z. Fang, X. Dai, H. Weng, Topological nodal line semimetals in the CaP 3 family of materials, Physical Review B. 95 (2017) 045136. doi:10.1103/PhysRevB.95.045136.

[28] G. Kresse, J. Furthmüller, Efficient iterative schemes for ab initio total-energy calculations using a plane-wave basis set, Phys. Rev. B. 54 (1996) 11169–11186. doi:10.1103/PhysRevB.54.11169.

[29] G. Kresse, J. Furthmüller, Efficiency of ab-initio total energy calculations for metals and semiconductors using a plane-wave basis set, Computational Materials Science. 6 (1996) 15–50. doi:10.1016/0927-0256(96)00008-0.

[30] J.P. Perdew, K. Burke, M. Ernzerhof, Generalized Gradient Approximation Made Simple, Phys. Rev. Lett. 77 (1996) 3865–3868. doi:10.1103/PhysRevLett.77.3865.

[31] P.E. Blöchl, Projector augmented-wave method, Phys. Rev. B. 50 (1994) 17953–17979. doi:10.1103/PhysRevB.50.17953.

[32] G. Kresse, D. Joubert, From ultrasoft pseudopotentials to the projector augmented-wave method, Phys. Rev. B. 59 (1999) 1758–1775. doi:10.1103/PhysRevB.59.1758.

[33] A.V. Krukau, O.A. Vydrov, A.F. Izmaylov, G.E. Scuseria, Influence of the exchange screening parameter on the performance of screened hybrid functionals, The Journal of Chemical Physics. 125 (2006) 224106. doi:10.1063/1.2404663.

[34] S. Grimme, J. Antony, S. Ehrlich, H. Krieg, A consistent and accurate *ab initio* parametrization of density functional dispersion correction (DFT-D) for the 94 elements H-Pu, The Journal of Chemical Physics. 132 (2010) 154104. doi:10.1063/1.3382344.

[35] A. Togo, F. Oba, I. Tanaka, First-principles calculations of the ferroelastic transition between rutile-type and ${\text{CaCl}}_{2}$-type ${\text{SiO}}_{2}$ at high pressures, Phys. Rev. B. 78 (2008) 134106. doi:10.1103/PhysRevB.78.134106.

[36] G.J. Martyna, M.L. Klein, M. Tuckerman, Nosé–Hoover chains: The canonical ensemble via continuous dynamics, The Journal of Chemical Physics. 97 (1992) 2635–2643. doi:10.1063/1.463940.

[37] K.-H. Xue, J.-H. Yuan, L.R.C. Fonseca, X.-S. Miao, Improved LDA-1/2 method for band structure calculations in covalent semiconductors, Computational Materials Science. 153 (2018) 493–505. doi:10.1016/j.commatsci.2018.06.036.

[38] L.G. Ferreira, M. Marques, L.K. Teles, Approximation to density functional theory for the calculation of band gaps of semiconductors, Phys. Rev. B. 78 (2008) 125116. doi:10.1103/PhysRevB.78.125116.

[39] J.-H. Yuan, Q. Chen, L.R. Fonseca, M. Xu, K. Xue, X. Miao, GGA-1/2 self-energy correction for accurate band structure calculations: the case of resistive switching oxides, Journal of Physics Communications. (2018) 105065. doi:10.1088/2399-6528/aade7e.

[40] C. Kamal, M. Ezawa, Arsenene: Two-dimensional buckled and puckered honeycomb arsenic systems, Physical Review B. 91 (2015). doi:10.1103/PhysRevB.91.085423.

[41] W. Wang, S. Dai, X. Li, J. Yang, D.J. Srolovitz, Q. Zheng, Measurement of the cleavage

energy of graphite, Nature Communications. 6 (2015) 7853.

[42] Y.-Q. Song, J.-H. Yuan, L.-H. Li, M. Xu, J.-F. Wang, K.-H. Xue, X.-S. Miao, KTlO: a metal shrouded 2D semiconductor with high carrier mobility and tunable magnetism, Nanoscale. (2019) 1131-1139. doi:10.1039/C8NR08046A.

[43] A. Savin, R. Nesper, S. Wengert, T.F. Fässler, ELF: The Electron Localization Function, Angewandte Chemie International Edition in English. 36 (1997) 1808–1832. doi:10.1002/anie.199718081.

[44] A.D. Becke, K.E. Edgecombe, A simple measure of electron localization in atomic and molecular systems, The Journal of Chemical Physics. 92 (1990) 5397–5403. doi:10.1063/1.458517.

[45] W. Tang, E. Sanville, G. Henkelman, A grid-based Bader analysis algorithm without lattice bias, Journal of Physics: Condensed Matter. 21 (2009) 084204. doi:10.1088/0953-8984/21/8/084204.

[46] G. Henkelman, A. Arnaldsson, H. Jónsson, A fast and robust algorithm for Bader decomposition of charge density, Computational Materials Science. 36 (2006) 354–360. doi:10.1016/j.commatsci.2005.04.010.

[47] E. Sanville, S.D. Kenny, R. Smith, G. Henkelman, Improved grid-based algorithm for Bader charge allocation, Journal of Computational Chemistry. 28 (2007) 899–908. doi:10.1002/jcc.20575.

[48] J. Bardeen, W. Shockley, Deformation Potentials and Mobilities in Non-Polar Crystals, Physical Review. 80 (1950) 72–80. doi:10.1103/PhysRev.80.72.

[49] H. Lang, S. Zhang, Z. Liu, Mobility anisotropy of two-dimensional semiconductors, Physical Review B. 94 (2016) 235306. doi:10.1103/PhysRevB.94.235306.

[50] M. Zhou, X. Chen, M. Li, A. Du, Widely tunable and anisotropic charge carrier mobility in monolayer tin(Ⅱ) selenide using biaxial strain: a first-principles study, Journal of Materials Chemistry C. 5 (2017) 1247–1254. doi:10.1039/C6TC04692D.

[51] C.G. Van de Walle, J. Neugebauer, First-principles calculations for defects and impurities: Applications to III-nitrides, Journal of Applied Physics. 95 (2004) 3851–3879. doi:10.1063/1.1682673.

Supporting Information for

# BaAs$_3$: A narrow gap 2D semiconductor with vacancy-induced semiconductor-metal transition

Ping Tang,[1+] Jun-Hui Yuan,[1+] Ya-Qian Song,[1] Ming Xu,[1] Kan-Hao Xue,[1,2*] Xiang-Shui Miao[1]

[1] Wuhan National Research Center for Optoelectronics, School of Optical and Electronic Information, Huazhong University of Science and Technology, Wuhan 430074, China

[2] Univ. Grenoble Alpes, Univ. Savoie Mont Blanc, CNRS, Grenoble INP, IMEP-LAHC, 38000 Grenoble, France

[+]The authors P. Tang and J.-H. Yuan contributed equally to this work.

**Corresponding Author**

*E-mail: xkh@hust.edu.cn (K.-H. Xue)

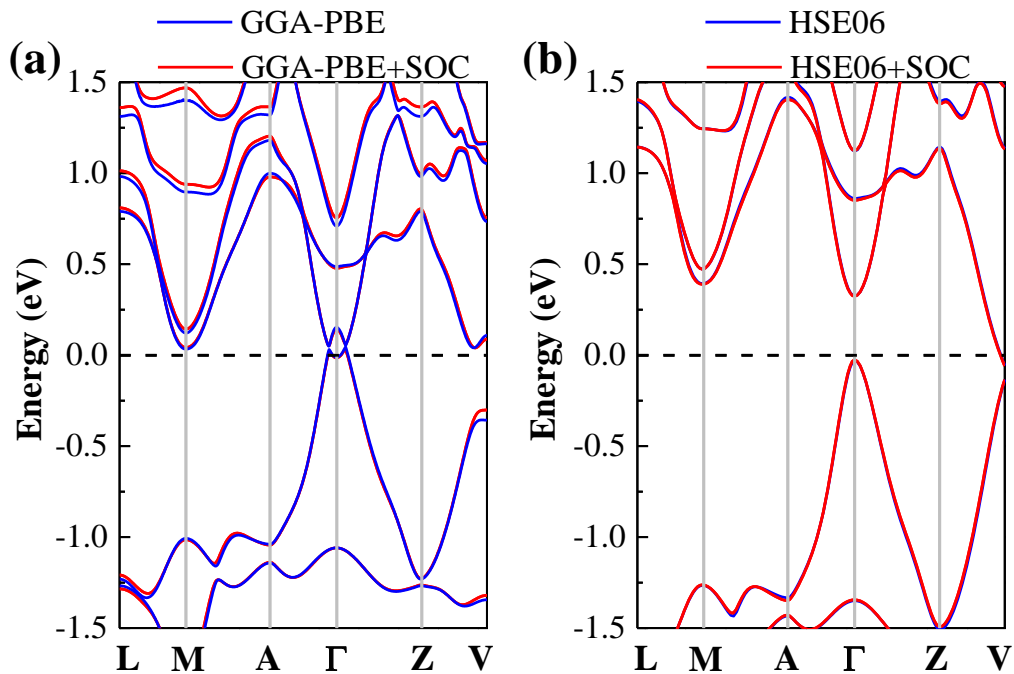

**Figure S1.** (a) Band structures of bulk BaAs$_3$ calculated using the PBE functional, either with or without considering the effect of spin orbit coupling (SOC). (b) Band structures of bulk BaAs$_3$ calculated using the screened exchange HSE06 hybrid functional, either with or without considering the effect of SOC.

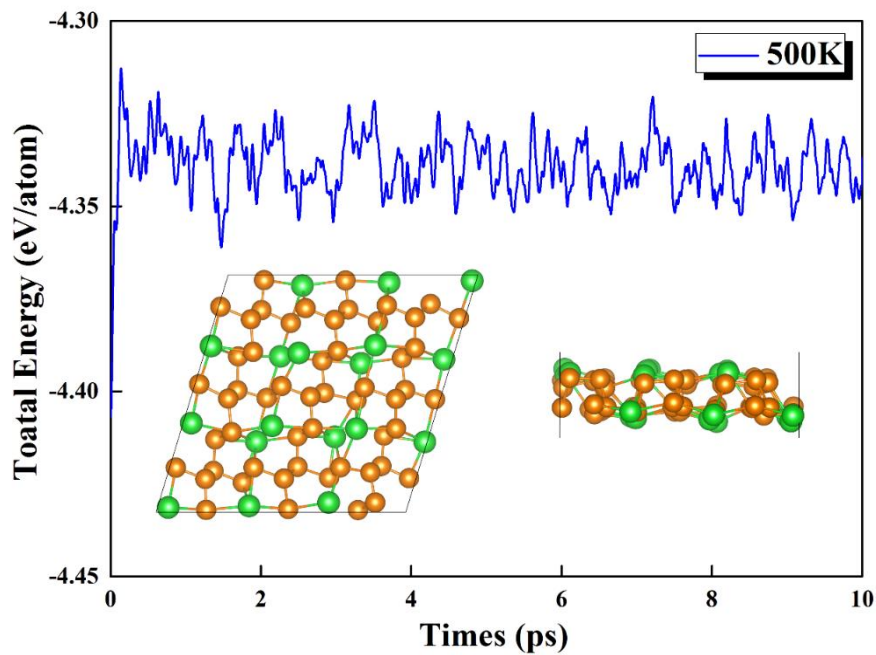

**Figure S2.** Top and side views of a snapshot from the molecules dynamics simulation for the ML BaAs$_3$. The variation of total energy is recorded and demonstrated over the simulation time of 10 *ps*, while the temperature was 500 K.

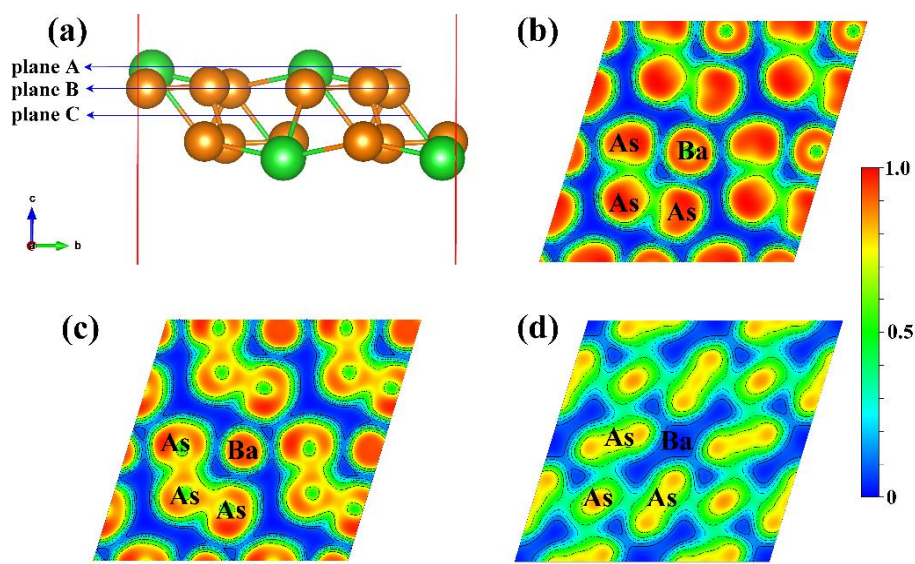

**Figure S3.** (a) Side view of ML BaAs$_3$ with planes A/B/C marked. Electron localization functions (ELFs) of: (b) plane A, (c) plane B, and (d) plane C, respectively.

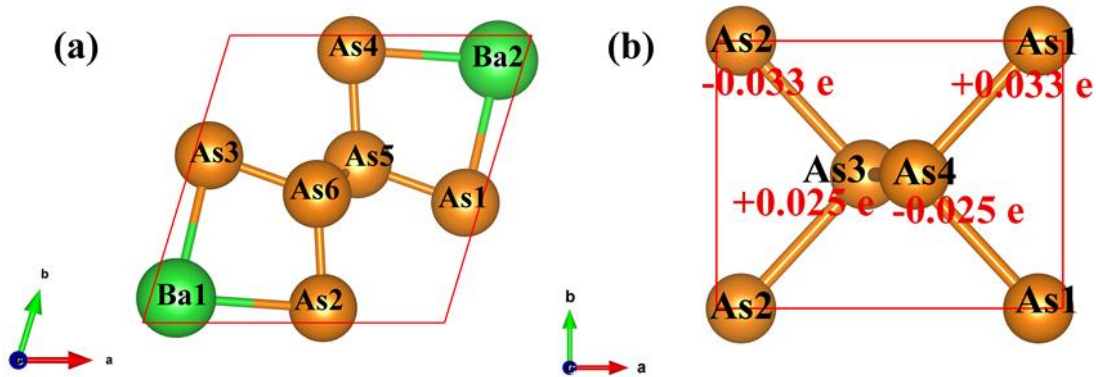

**Figure S4.** (a) Top view of ML BaAs$_3$ with all atoms marked by element symbol and numbers. (b) Top view of puckered arsenene with Bader charge values marked.

**Table S1.** Charge transfer between Ba and As atoms in ML BaAs$_3$ according to the Bader gauge. Here "+" and "-" represent a loss and a gain of electrons, respectively.

| | Ba1 | Ba2 | As1 | As2 | As3 | As4 | As5 | As6 |
|---|---|---|---|---|---|---|---|---|
| Charge (\|e\|) | +1.3186 | +1.3186 | -0.5614 | -0.5723 | -0.5680 | -0.5720 | -0.1794 | -0.1843 |

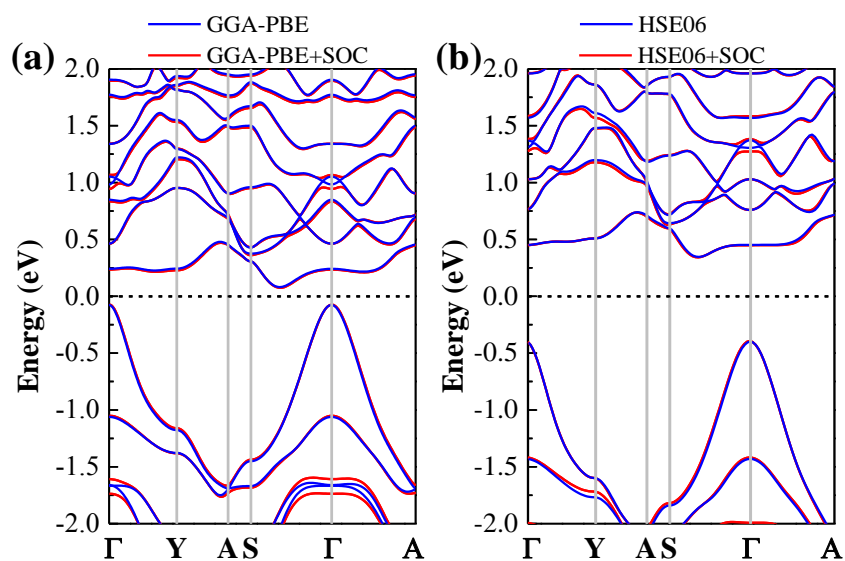

**Figure S5.** (a) Band structures of ML BaAs$_3$ calculated using the PBE functional, either with or without considering the effect of spin orbit coupling (SOC). (b) Band structures of ML BaAs$_3$ calculated using the screened exchange HSE06 hybrid functional, either with or without considering the effect of SOC.

S3

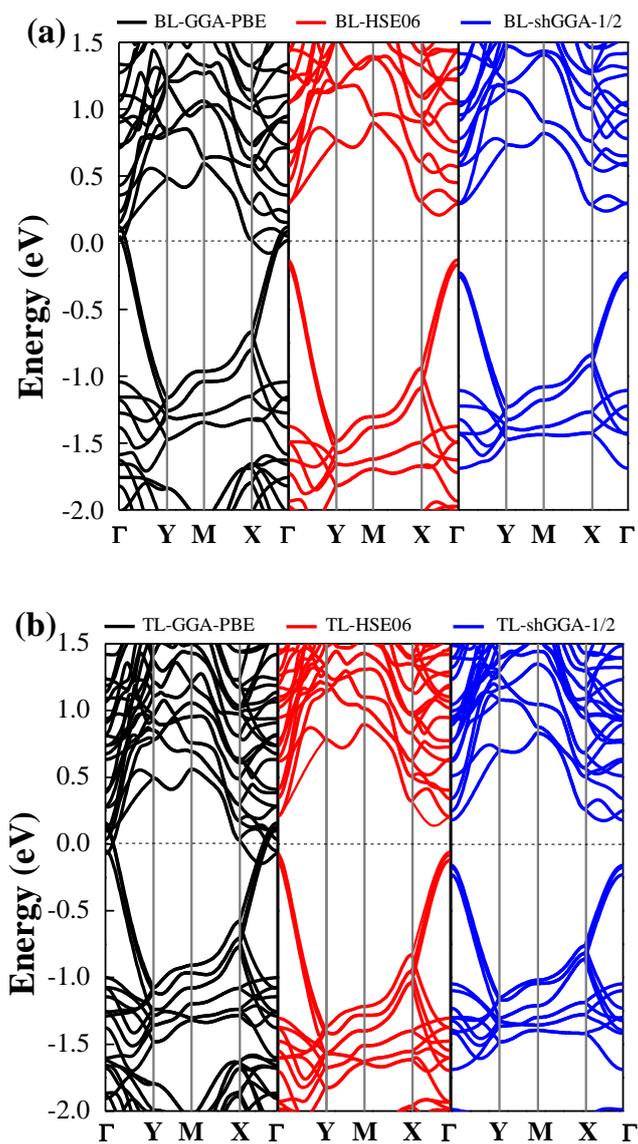

**Figure S6.** Electronic band structures of 2D BaAs$_3$ with varying number of layers, calculated using GGA-PBE, HSE06 hybrid functional and the shGGA-1/2 method: (a) bilayer (BL for short); (b) trilayer (TL for short).

**Table S2.** The optimized lattice constants (*a*/*b*/*c*) in ML BaAs$_3$, BL, TL and bulk, respectively, in comparison to the experimental bulk values (Exp.).

| Lattice constant | ML | BL | TL | Bulk | Exp.[1] |
|---|---|---|---|---|---|
| *a*/Å | 6.51 | 10.27 | 10.26 | 10.22 | 11.16 |
| *b*/Å | 6.25 | 7.80 | 7.81 | 7.82 | 7.76 |
| *c*/Å | -- | -- | -- | 6.06 | 6.01 |
| α/ ° | 89.76 | 90 | 90 | 90 | 90 |
| β/ ° | 89.76 | 90.07 | 106.40 | 113.70 | 113.55 |
| γ/ ° | 73.19 | 90 | 90 | 90 | 90 |

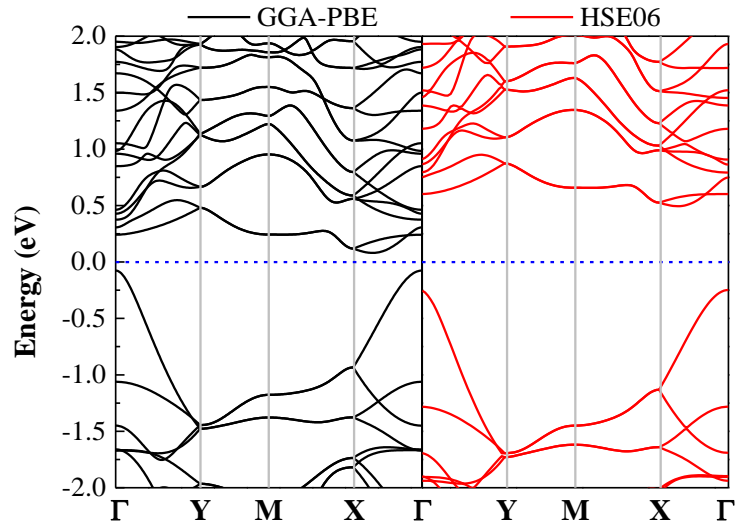

**Figure S7.** Electronic band structures of ML BaAs$_3$ in an orthogonal supercell, calculated using both GGA-PBE and the screened exchange HSE06 hybrid functional.

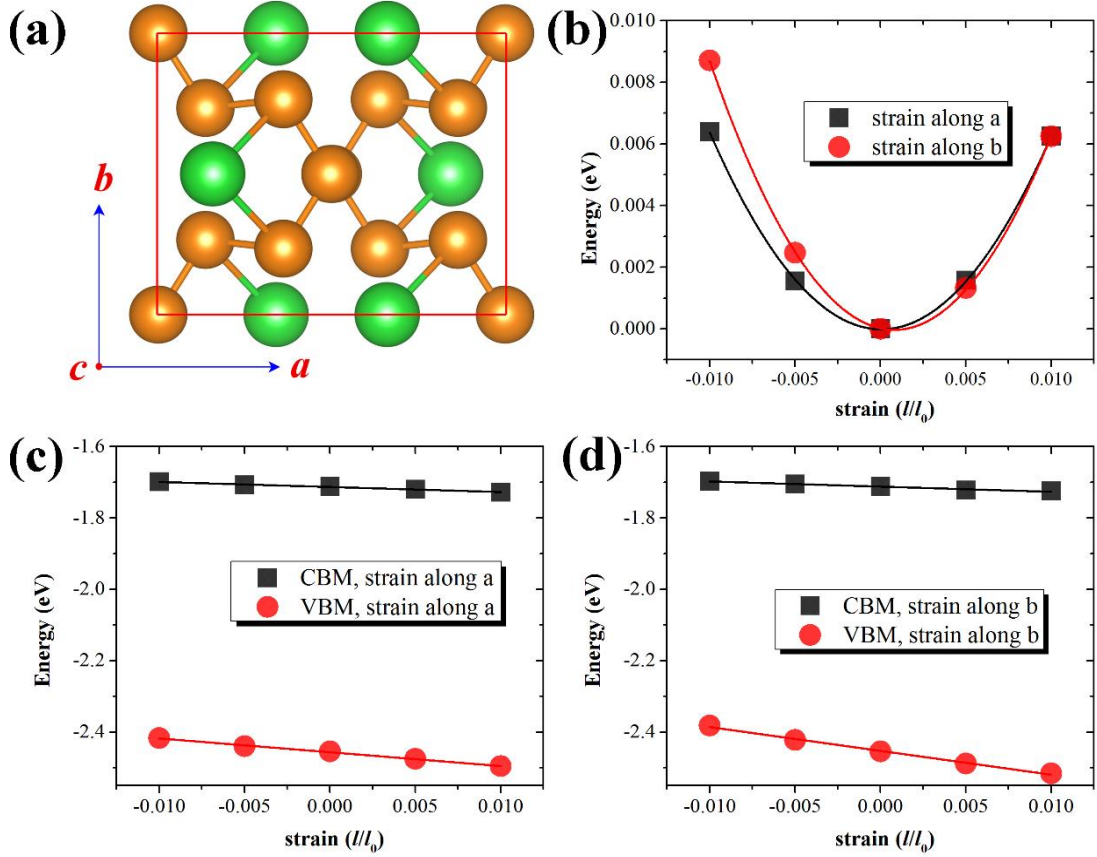

**Figure S8.** (a) The orthogonal supercell of ML BaAs$_3$. (b) The relation between total energy and the applied strain $\delta$ along the $a$ and $b$ directions of ML BaAs$_3$. The quadratic data fitting gives the in-plane stiffness of 2D structures. Black and red curves show the in-plane stiffness along the $a$ and $b$ directions of ML BaAs$_3$, respectively. (c) The shift of VBM and CBM for ML BaAs$_3$ with respect to the vacuum energy, as a function of the applied strain along the $a$ direction. (d) The shift of VBM and CBM for ML BaAs$_3$ with respect to the vacuum energy, as a function of the applied strain along the $b$ direction. The linear fit of the data yields the deformation potential constant. All the calculations were based on the HSE06 functional.

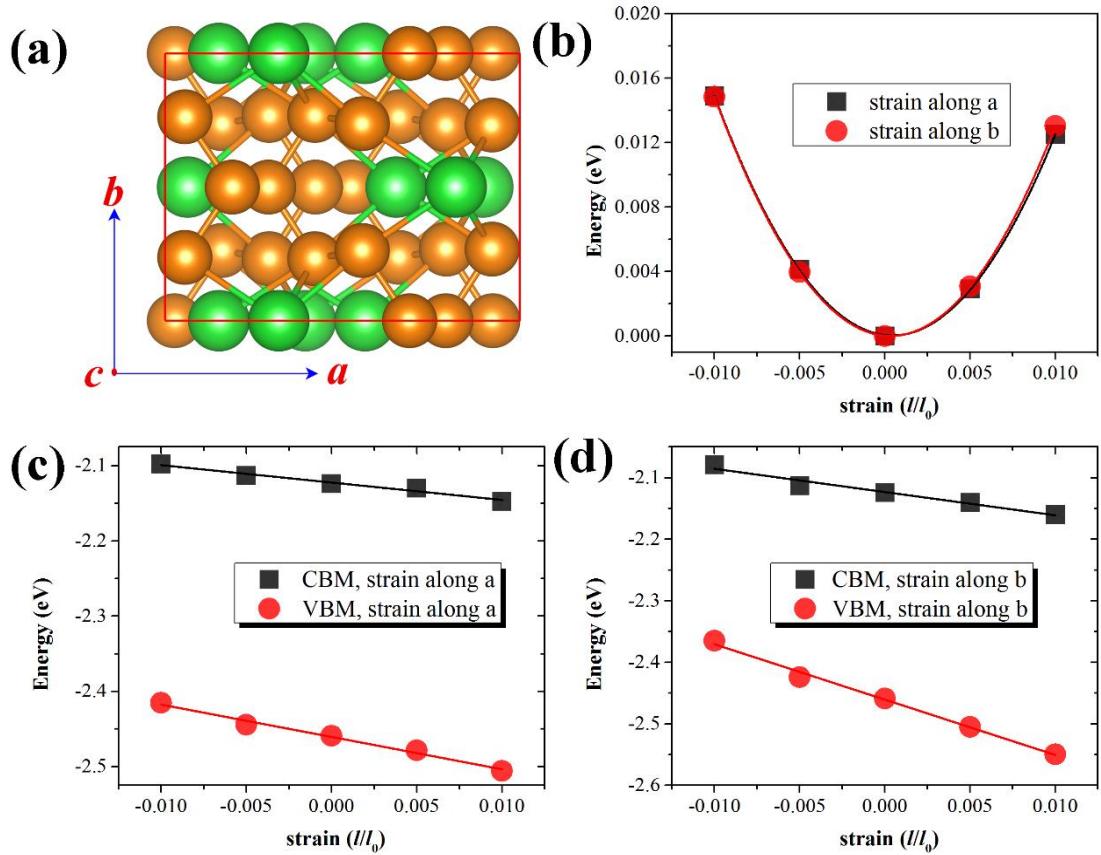

**Figure S9.** (a) The orthogonal supercell of bilayer BaAs$_3$. (b) The relation between total energy and the applied strain $\delta$ along the *a* and *b* directions of bilayer BaAs$_3$. The quadratic data fitting gives the in-plane stiffness of 2D structures. Black and red curves show the in-plane stiffness along the *a* and *b* directions of bilayer BaAs$_3$, respectively. (c) The shift of VBM and CBM for bilayer BaAs$_3$ with respect to the vacuum energy, as a function of the applied strain along the *a* direction. (d) The shift of VBM and CBM for bilayer BaAs$_3$ with respect to the vacuum energy, as a function of the applied strain along the *b* direction. The linear fit of the data yields the deformation potential constant. All the calculations were based on the HSE06 functional.

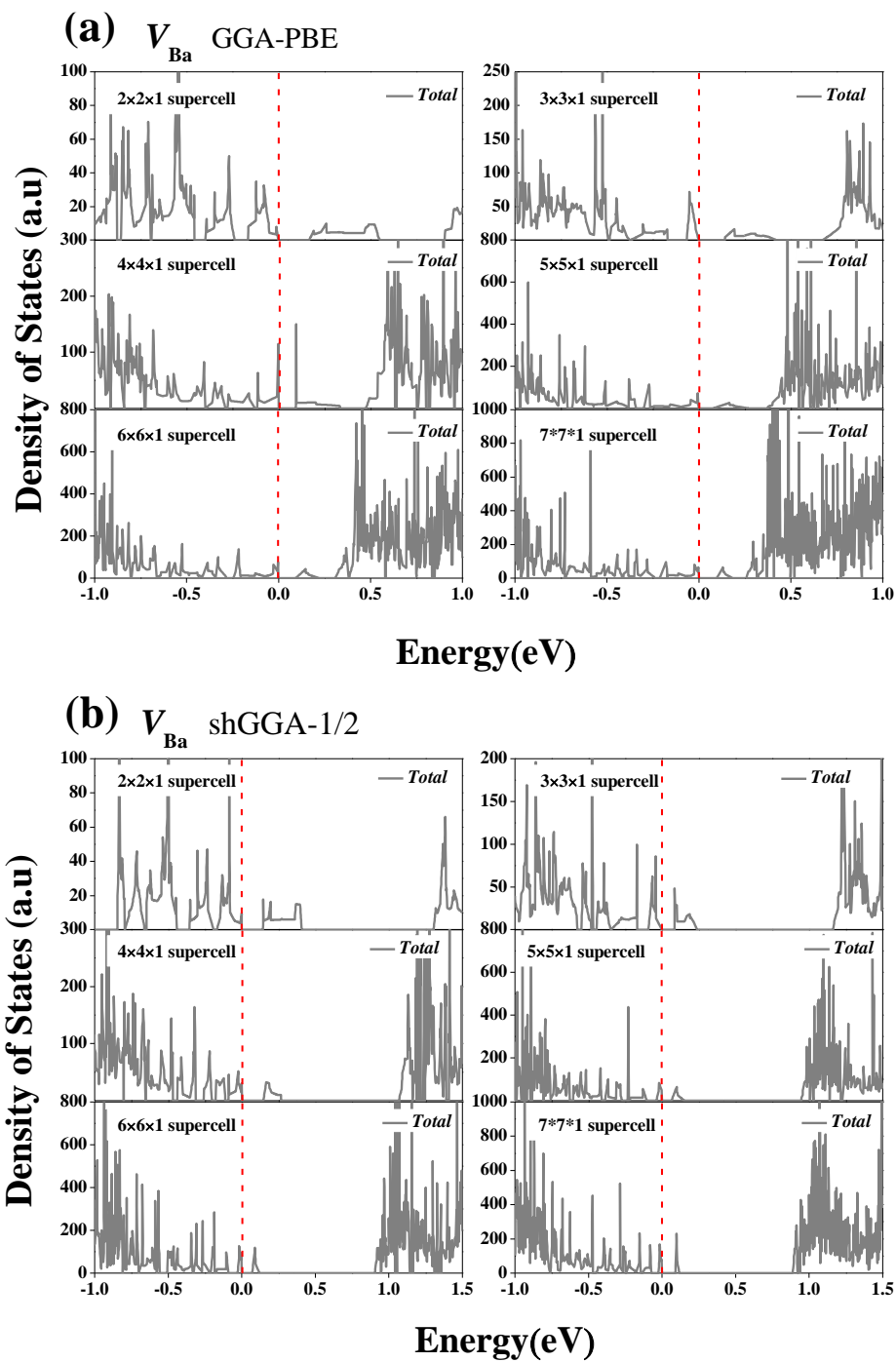

**Figure S10.** Partial DOS of $V_{Ba}$-containing supercells from $2\times2\times1$ to $7\times7\times1$, calculated using (a) the GGA-PBE functional; (b) the shGGA-1/2 method.

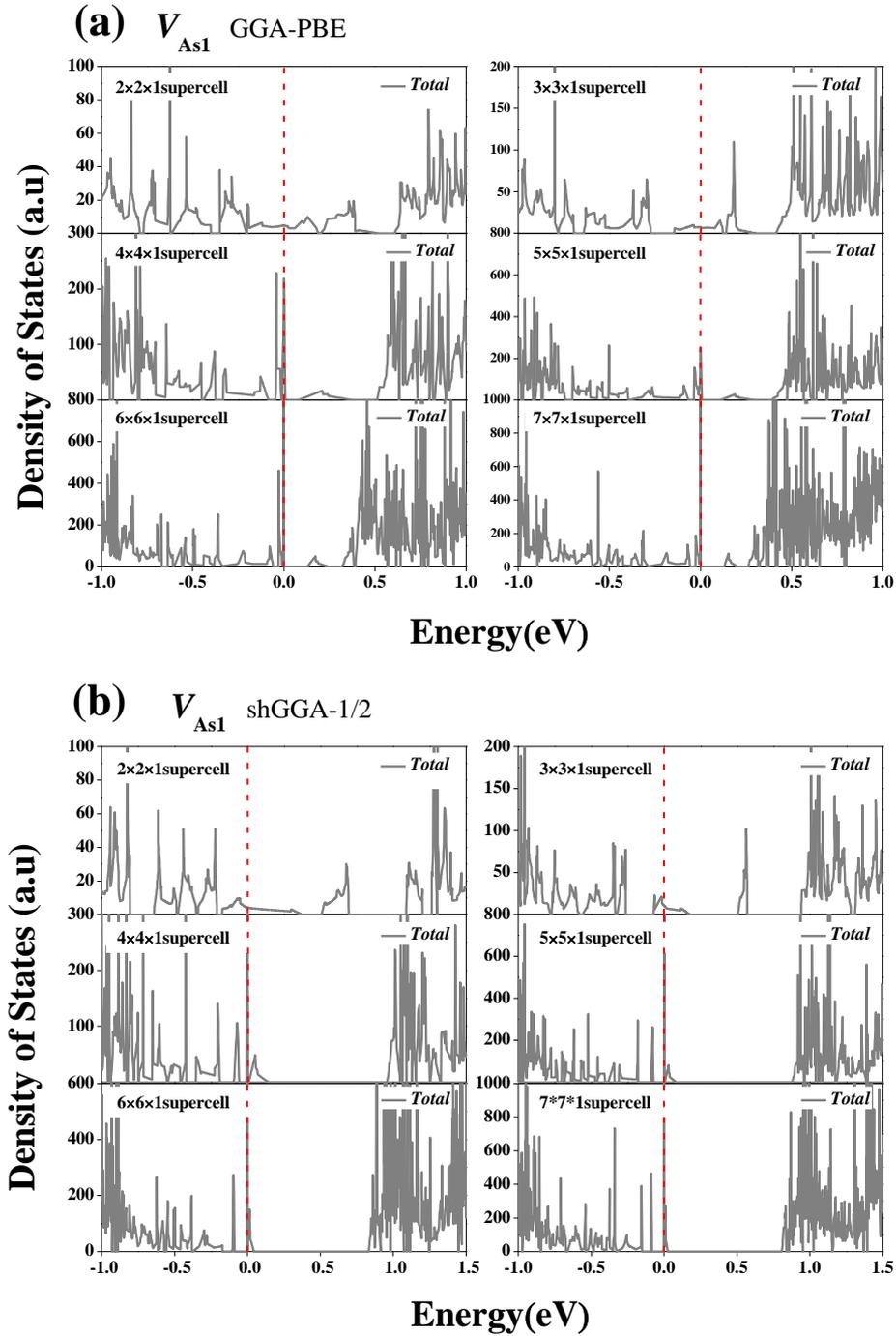

**Figure S11.** Partial DOS of $V_{As1}$-containing supercells from $2\times2\times1$ to $7\times7\times1$, calculated using (a) the GGA-PBE functional; (b) the shGGA-1/2 method.

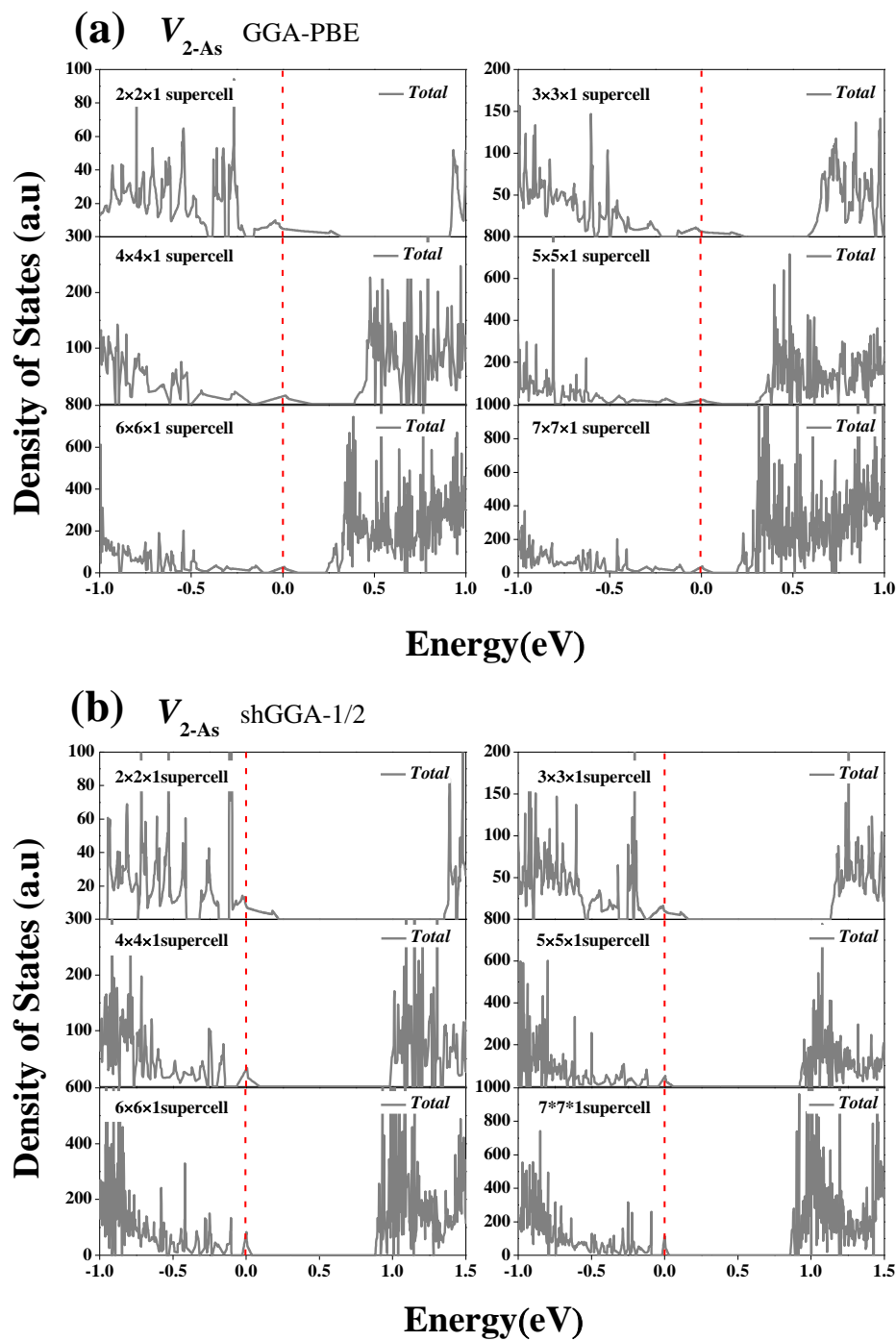

**Figure S12.** Partial DOS of $V_{As2}$-containing supercells from $2\times2\times1$ to $7\times7\times1$, calculated using (a) the GGA-PBE functional; (b) the shGGA-1/2 method.

**REFRENCES**

(1) Bauhofer, W.; Wittmann, M.; Schnering, H. G. v. Structure, Electrical and Magnetic Properties of CaAs3, SrAs3, BaAs3 and EuAs3. *Journal of Physics and Chemistry of Solids* **1981**, *42* (8), 687–695. https://doi.org/10.1016/0022-3697(81)90122-0.